# Size, Shape and Phase Control in Ultrathin CdSe Nanosheets

*Frauke Gerdes,[†] Cristina Navío,[‡] Beatriz H. Juárez,[‡, §] and Christian Klinke[†,*]*

† Institute of Physical Chemistry, University of Hamburg, Grindelallee 117, 20146 Hamburg, Germany.

‡ Madrid Institute of Advanced Studies in Nanoscience, IMDEA Nanoscience, Faraday 9, Cantoblanco, 28049 Madrid, Spain.

§ Applied Physical-Chemistry Department, Universidad Autónoma de Madrid, Cantoblanco, 28049 Madrid, Spain.

**Abstract**

Ultrathin two-dimensional nanosheets raise a rapidly increasing interest due to their unique dimensionality-dependent properties. Most of the two-dimensional materials are obtained by exfoliation of layered bulk materials or are grown on substrates by vapor deposition methods. To produce free-standing nanosheets, solution-based colloidal methods are emerging as promising routes. In this work, we demonstrate ultrathin CdSe nanosheets with controllable size, shape and phase. The key of our approach is the use of halogenated alkanes as additives in a hot-injection synthesis. Increasing concentrations of bromoalkanes can tune the shape from sexangular to quadrangular to triangular and the phase from zinc blende to wurtzite. Geometry and crystal structure evolution of the nanosheets take place in the presence of halide ions, acting as cadmium complexing agents and as surface X-type ligands, according to mass spectrometry and X-ray photoelectron spectroscopies. Our experimental findings show that the degree of these changes depends on the molecular structure of the halogen alkanes and the type of halogen atom.

Ultrathin two-dimensional (2D) nanocrystals have attracted increasing research interests due to their unique properties.[1–3] These properties are based on the fact that such nanosheets (NSs) have lateral dimensions of at least 100 nm, but are only a few atomic monolayers thick, which creates a strong 1D quantum confinement. Since the experimental discovery of graphene,[4,5] there are fundamental research efforts to synthesize ultrathin NSs of various materials, like metals, oxides and chalcogenides.[6–9] Most of these NSs were obtained by exfoliation of layered bulk materials or were grown on substrates by molecular beam epitaxy and chemical vapor deposition.[10–12] To produce large amounts of free-standing NSs and to control their shape and lateral dimensions, colloidal chemistry is a promising approach. Though, the synthesis of 2D colloidal semiconductor nanocrystals is limited to a few examples[13–16] because of the difficult growth control in specific directions, in contrast to the well-documented 0D and 1D systems.[17–19] One of the most investigated 2D colloidal systems are CdSe nanoplatelets[20–22] and nanoribbons,[23–25] which have been shown to possess particular optical properties.[26–28] CdSe nanoplatelets are formed by continuous lateral growth of small nanocrystals, which results in structures with the cubic phase zinc blende (ZB),[21] while CdSe nanoribbons are generated by a soft-template method and show hexagonal phase wurtzite (WZ) crystal structure.[24] Due to the different colloidal synthesis strategies and conditions, the properties of both systems are difficult to compare.

Here, we show that the size, shape and phase of ultrathin colloidal CdSe NSs can be controlled by the addition of halogenated compounds to the synthesis. For example, with increasing amounts of 1-bromoheptane (Br-Hep) the shape can be tuned from sexangular to quadrangular to triangular and the phase changes from ZB to WZ. We systematically investigate the influence of the chemical structure of the bromoalkanes as well as the type of the halogen atom in the additive to understand the role of halogen compounds and to investigate the growth mechanism of the CdSe NSs.

In a first series of experiments the amount of Br-Hep has been varied. Transmission electron microscopy (TEM) images (Figure 1) show the shape evolution of the CdSe NSs upon addition of Br-Hep. CdSe NSs without Br-Hep (Figure 1A) have a lateral width of $28 \pm 5$ nm and a lateral length of $69 \pm 8$ nm. The thickness of the NSs is $2.0 \pm 0.4$ nm, directly measured by TEM on NSs which stand vertical on the TEM grid. Adding and increasing the amount of Br-Hep, the shape of the CdSe NSs changes from sexangular (no Br-Hep, Figure 1A) to quadrangular (0.0625, 0.125 and 0.25 mmol Br-Hep, Figure 1B, 1C and 1D) to a mixture of different forms (0.5 mmol Br-Hep, Figure 1E), and to triangular (1 and 2 mmol Br-Hep, Figure 1F and 1G) ones. The triangular CdSe NSs (Figure 1G) with truncated tips show a length of $65 \pm 9$ nm and a thickness of $2.3 \pm 0.3$ nm. Due to the effective instrument resolution of the TEM the actual thickness of the CdSe NSs is probably less than the determined one, which is in agreement with the results of the Dubertret group.[29] At higher concentrations of Br-Hep, CdSe NSs are first formed, indicated by an orange-yellow, turbid solution, and then, with continued reaction time the reaction solution becomes nearly colorless and a yellowish reaction product is isolated (4 mmol Br-Hep, Figure 1H).

X-ray diffraction (XRD) was performed to determine the crystal structure of the CdSe NSs. The XRD patterns (Figure 2) indicate that a change in the phase from ZB to WZ occurred as the shape of the CdSe NSs varied from sexangular to triangular. Without Br-Hep on the synthesis, the characteristic diffraction peaks correspond to the cubic ZB structure (ICPDS 00-019-0191). Increasing the amount of Br-Hep, the cubic (400) peak disappears and the characteristic peak of hexagonal (10-12) and (10-13) planes corresponding to the WZ structure became discernible, revealing the phase changes from ZB to WZ with higher amounts of Br-Hep. When 2 mmol Br-Hep was used, the XRD pattern displays all characteristic peaks of the bulk WZ-CdSe pattern (ICPDS 00-008-0459). The selected area electron diffraction (SAED) pattern (see Supporting Information, Figure S1) supports the findings of phase change from ZB to WZ with increasing amounts of Br-Hep. The XRD pattern of the sample synthesized with 4 mmol Br-Hep shows a strong texture effect and can be assigned to bulk $CdBr_2$ (ICPDS 00-010-0438). This observation indicates that bromide anions ($Br^-$) must be formed *in-situ* and that they might be responsible for the shape and phase change of the CdSe NSs. This notion will be further supported in the following discussion.

The XRD patterns were also analyzed to identify the growth direction of the ZB- and WZ-CdSe NSs. Peng and co-workers showed that the width of the peaks is correlated with the growth directions for ZB-CdSe nanoplatelets.[30] A broad peak is assigned to short axis directions, while a sharp peak is related to lateral directions. For the ZB-CdSe NSs, the short axes are along the [001] and <111> direction, whereas the [001] direction corresponds to the sheet thickness. The (220) peak shows a superposition of one broad and one sharp peak. This observation is explained by the two-dimensionality of the structure. Due to the anisotropic structure, the XRD peaks which correspond to the lattice planes in different axis directions have different positions and widths and do not overlapped completely. As a result, the planes perpendicular to the short axis have a narrower peak than the ones parallel to the short axis.[30] Thus, the [100], the [010], and the [110] direction can be determined as lateral growth direction. Due to the asymmetry of the WZ phase, the growth direction determination suggested by the Peng group cannot be adopted for the growth direction in the WZ-CdSe NSs. However, Peng and co-workers simulated XRD patterns of WZ-CdSe 2D nanocrystals with different orientations, in which they assume that the [0001] direction is one of the lateral direction due to the easy growth along this axis.[31] If the thickness is in [10-10] direction and the lateral growth is along the [0001] and [11-20] axis, the calculated XRD shows sharp (002) and (110) peaks with increased intensity, while the (100) peak is broadened and less intense. For the WZ-CdSe NSs, we observed a very similar XRD pattern with an additional sharp peak for the (112) lattice planes. Supported by HRTEM (see Supporting Information, Figure S2A), the lateral growth is along the <0001>, <11-20> and <11-22> directions. The higher intensity of the (002) XRD peak relative to the other sharp peaks substantiates the main growth in <0001> direction. The scheme of a truncated triangular WZ-CdSe NSs shows that the lateral facets are either Cd-rich (large) or Se-rich (small), while the planar {10-10} facet exhibit a mixed, arm-chaired composition (see Supporting Information, Figure S2C and S2D).

The optical properties of the CdSe NSs were characterized by UV-vis absorption and photoluminescence (PL) spectroscopy. The absorption spectrum of the ZB-CdSe NSs (Figure 3A) shows the first heavy hole (hh)-exciton transition[28,32] at 463 nm. Also the first light hole (lh)-exciton transition at 434 nm as well as the second exciton transitions at 393 nm and 370 nm are observed. The doublets in the absorption spectra support the 2D shape of the CdSe NSs by using the model of infinite potential quantum wells.[33] The PL spectrum of the ZB-CdSe NSs (Figure 3A) shows a narrow emission band with a full width at half maximum (FWHM) of 8.9 nm at 465 nm, thus confirming the homogeneous thickness of the CdSe NSs. The broad band at higher wavelength is explained by trap states in the NSs. The absorption spectrum of the WZ-CdSe NSs (Figure 3B) shows a weak first hh-exciton transition at 515 nm and the second prominent exciton transition at 466 nm and 431 nm, which may be generated by two NSs populations with different thickness. Accordingly, in the PL spectrum (Figure 3B) a strong emission band with a FHWM of 22.7 nm at 513 nm and a weaker emission band with a FHWM of 13.3 nm at 464 nm are observed. The found absorption and emission bands for ZB-CdSe NSs are comparable to those for the CdSe nanoplatelets by the Dubertret group.[20–22] The sharp absorption and emission features of the CdSe NSs clearly demonstrate that the origin must be structures which are strongly in confinement. We also measured the quantum yield (QY) in respect to the standard quinine sulfate. The ZB-CdSe

NSs have a QY of 0.5 %, while the QY of the WZ-CdSe NSs is 1 %. These QYs are probably low due to scattering due to the large lateral NS size and to trap states at the surface.

It is important to mention that according to literature, the formation of the thermodynamically stable WZ is promoted by phosphonate and amine ligands, while the formation of the metastable ZB is favored by carboxylate ligands.[34] Besides, the size and shape dependent phase transformations observed for ZnS and CdS nanocrystals,[35–37] as well as the effect of the new ion-ligand coordination during nucleation and growth [34,38,39] induce the transition from ZB structure to WZ structure.

To better understand the mechanism of the phase and shape control of the CdSe NSs, we performed a series of experiments with different halogen compounds. If the additive affects the kinetics of the nucleation and growth process, different halogen compounds should be able to modify the formation of the NSs. Firstly, we investigated the effects of the chemical structure of the bromoalkane. For this, bromoalkanes with different chain lengths ($C_2 - C_{14}$) and numbers of bromine atoms ($Br_1 - Br_4$) were used in different amounts ($0.0625 - 4$ mmol) after 10 min of reaction (see Supporting Information, Table S1 and Figure S4). All additives were directly injected after the vacuum step at 80 °C.

In all cases, triangular WZ-CdSe NSs were obtained but the minimum required amount of bromoalkane differs notably. Moreover, the size and the angle of the tips of the NSs vary. The series of bromoalkanes with different chain length shows that the required concentration to produce WZ-CdSe NSs decrease with increasing chain length. For example, using 1-bromoethane ($Br_1$-$C_2$) a concentration of 1 mmol is necessary (Figure 4A), while only 0.5 mmol of 1-bromotetradecane ($Br_1$-$C_{14}$) is required (Figure 4B). Moreover, it was observed that with increasing chain length the truncated triangles are better defined. Increasing the number of bromine atoms at the bromoalkane leads also to smaller required amounts of the alkane to form WZ-CdSe NSs. Furthermore, the angle of the tips of the triangles decreases from 120 ° to 60 ° and the size of the triangles increases with a higher number of bromine atoms. For example, the use of 1,10-dibromodecane ($Br_2$-$C_{10}$) resulted in NSs with a size up to 150 nm (Figure 4C), while using 1,1,2,2-tetrabromoethane ($Br_4$-$C_2$) triangular NSs with a size up to 200 nm were synthesized (Figure 4D) at comparable reaction times. $Br_4$-$C_2$ is a special case because it is the only bromoalkane, which never produced NSs with ZB structure. Further, it does not show a shape change from sexangular to quadrangular but triangular NSs are found for all concentrations (see Supporting Information, Figure S4).

To explain these observations it is important to mention that in previous studies we found that halogenated compounds have an important influence on the shape of semiconductor nanocrystals.[40,41] We observed that chlorinated additives are indispensable for the formation of two-dimensional PbS nanosheets.[40] We also observed that halogenated additives lead to a shape evolution from WZ-CdSe nanorods into pyramidal-shape nanoparticles.[41] In this case the additives work as X-type (ionic) ligands and coordinate to Cd-rich facets, leading to a modified chemical composition of the ligand sphere. We also evidenced that the active species promoting the shape transformation were halide ions produced *in-situ*, that influence both the nucleation and growth of the nanostructures.[42] In the present work, the addition of halogenated compounds leads to the formation of CdSe NSs with tunable size, shape and phase. Thus, it is clear that they are the key to control the CdSe NSs.

To determine the influence of the bromoalkane additive in the reaction and acquire knowledge about the mechanism of the reaction, we firstly performed electron ionization (EI) mass spectrometry (MS). For MS measurements, samples obtained with Br-Hep were taken from the reaction mixture after the vacuum step (before the Br-Hep injection) and directly before the injection of the TOPSe (10 min after Br-Hep injection). In the EI spectra of the samples (see Supporting Information, Figure S5), the dominated peaks at *m/z* 77.1, 141.1 and 170.1 belong to the solvent diphenyl ether (see Supporting Information, Figure S6). The detailed EI spectrum of the sample without Br-Hep (Figure 5A top) shows the crucial peak at $m/z$ 231.2. This peak can be assigned to the cadmium acetate compound [(CH$_3$COO)-Cd-(OOCCH$_3$)] and may correspond to a singly charged species. The important peaks of the EI spectrum of the sample with Br-Hep (Figure 5A bottom) are at $m/z$ 248.1, 249.1 and 250.1, which may be attributed to a cadmium acetate bromide complex [Br-Cd-(OOCCH$_3$)], evidencing the formation of Br$^-$ during the reaction. The three peaks must correspond to the deprotonated species [M-1H]$^+$, [M-2H]$^+$, and [M-3H]$^+$. Due to the lower concentration of oleate in comparison to the acetate and the fragmentation in EI mass spectrometry, peaks related to cadmium oleate ($m/z$ 675) or cadmium oleate bromide complex are not observed ($m/z$ 474). The McLafferty rearrangement of the cadmium oleate complex leads to the main fragment ion peak $m/z$ 232, which corresponds to the cadmium acetate complex. The spectrum also shows that the cadmium acetate compound still remains in the solution after the addition of Br-Hep, suggesting that only a part of the bromine additive bonds to cadmium.

Furthermore, to determine the active species of the bromine additive which bind to the CdSe NSs surface, we performed X-ray photoelectron spectroscopy (XPS). For the XPS measurements, the pure CdSe NSs solution was drop-casted on a highly oriented pyrolytic graphite (HOPG) substrate. HOPG was taken as binding energy reference with the C *1s* core level at 284.5 eV. The sample was fixed to a Mo sample plate with C scotch tape. The Cd *3d* region in the XPS spectra suggests a contribution of Cd bound to Br when Br-Hep was added to the synthesis (see Supporting Information, Figure S7). The XPS spectrum of the WZ-CdSe NSs for the Br *3d* area (Figure 5B) can be fitted with contributions of two different components, indicating different chemical Br environments. According to previous works, the peak at 70.0 eV can be assigned to covalently bonded bromine atoms in Br-Hep, while the smaller peak at 66.6 eV can be related to ionic bonded bromine atoms (bromide, Br⁻), as previously reported for carbon nanotubes.[43–46]

Thus, from MS and XPS we can conclude that the bromine additive has two active species, Br-Hep and ionic Br⁻. The presence of two active species on the surface of the NSs suggests that both are important. The presence of Br⁻ suggests that the results shown in Figure 4 might be explained according to the reactivity of the haloalkane in nucleophilic substitution and/or elimination reactions.[47] As evidenced in Figures 4A (1-bromoethane) and 4B (1-bromotetradecane), the longer the alkyl chain in the added primary haloalkane, the smaller is the concentration required to obtain triangular NSs that evolve from quadrangular ZB NSs. This scenario may be due to a higher reactivity of 1-bromotetradecane to release Br⁻, although it should be taking into account that, according to XPS, not only Br⁻ but also the molecule are active species capping the NSs surface and thus, alternative mechanisms may take place. The effect that the amount of bromine atoms in the molecule produces in the reaction is exemplified in Figure 4C, where only 0.25 mmol of the primary haloalkane is necessary. The effect of a secondary haloalkane can be observed in Figure 4D. In this case due to the higher steric hindrance of the carbon atom and the high –I effect of the bromine atoms, the bromoalkane is less stabilized and has consequently a higher reactivity. Thus, triangular NSs are obtained from concentration as low as 0.0625 mmol.

To further investigate the influence of the type of halogen atom and to control size, shape and phase of the NSs, we utilized chemical analogs of Br-Hep, namely 1-chloroheptane (Cl-Hep) and 1-iodoheptane (I-Hep) in equimolar amounts (0.0625 – 4 mmol). TEM images (Figure 6) and XRD diffractograms (see Supporting Information, Figure S8 and S9) of the Cl-Hep and I-Hep samples show that these haloalkanes influence the formation and final shape of the CdSe NSs in a different way. Using low amounts of Cl-Hep the shape of NSs change from sexangular (0.0625 mmol, Figure 6A) to quadrangular (0.125 mmol, Figure 6B), while the crystal phase is ZB (see Supporting Information, Figure S8). In comparison to Br-Hep, increasing amounts of Cl-Hep lead to no further changes in shape or phase of the NSs. In contrast, adding and increasing the amount of I-Hep leads to a shape change of the synthesized NSs from sexangular (no additive, Figure 1A) to rolled-up quadrangular (0.0625 and 0.125 mmol, Figure 6F and 6G) to triangular (0.5 mmol, Figure 6H) to hexagonal (2 and 4 mmol, Figure 6I and 6J). The XRD diffractograms of the I-Hep samples (see Supporting Information, Figure S9) show that first, the crystal phase changed from ZB-CdSe (0.0625 and 0.125 mmol) to WZ-CdSe (0.5 mmol), and then the material changed from CdSe to $CdI_2$ (2 and 4 mmol), similar to the observations using Br-Hep.

These different results shown in Figure 6 might be once more explained by the nucleophilic character and leaving group tendency of the halogens in substitution and/or elimination reactions, which increase with heavier atoms.[47] Most probably this is induced by a Cd-bonded acetate acting as a nucleophile or base. As a consequence, for similar initial concentrations of haloalkanes, the concentration of the corresponding released halide ions ($I^-$, $Br^-$ and $Cl^-$) in the reaction mixture would be higher for I-Hep > Br-Hep > Cl-Hep. In other words, the carbon-halogen bond is weaker in C-I compared to C-Br and C-Cl. Small amounts of I-Hep seem to be sufficient to modify both shape and phase, while higher amounts are needed for Br-Hep and no modifications are evidenced for Cl-Hep at comparable reaction times, temperature and concentration. Based on this, we propose a mechanism for the impact of the bromine additive on the size, shape and phase of CdSe NSs with increasing amount of additive.

As evidenced by MS and XPS measurements in this work the bromine additive has two active species, Br-Hep and ionic $Br^-$. The *in-situ* formed $Br^-$ concentration depends on the amount of added Br-Hep to the synthesis. If the amount of bromoalkane and consequently the amount of $Br^-$ is low, only the shape of the ZB-CdSe NS changes. It is known that the shape of nanocrystals is determined during the growth stage by the different growth rates of the facets. Lim *et al.* showed by [1]H-NMR that the amount of oleate ligands at the surface of CdSe tetrapods decreased with the increase of halide ions and the replacement of oleate induces an anisotropic growth.[38] In addition, the XPS results reported here support that the bromine additive acts as surface ligand in the form of Br-Hep and ionic Br⁻. They partially replace the carboxylate ligands on the surface which harmonizes with the hard-soft acid-base (HSAB) concept, where soft bases bind more strongly to soft acids, so that the bromine additive is in the position to displace the carboxylates and form Cd-Br bonds. The new surface ligands must lead to a regulation of the growth rate by modulation of the relative surface energies of those facets due to its different binding affinities to bromine.

When the addition of Br-Hep and the resulting amount of Br⁻ is increased, according to our results, not only the shape but also the phase of the CdSe NSs varies. It is feasible that in this case, for higher Br⁻ concentrations, the nucleation stage might be also influenced. As evidenced by MS, the Br⁻ ions partially replace the acetate and form a new cadmium precursor, a cadmium acetate bromide complex [Br-Cd-(OOCCH$_3$)]. Due to the Cd-Br bond the activated complex formed upon reaction with Se–P(octyl)$_3$, ([Br-Cd-Se–P(octyl)$_3$]$^+$), must be less reactive than the previously evidenced general complex [RCOO-M-E–P(octyl)$_3$]$^+$, where M stands for a metal and E stands for a chalcogen, that forms when synthesizing chalcogenide nanocrystals by using metal carboxylates ((RCOO)$_2$M) and tri-*n*-octyl phosphine chalcogenides ((octyl)$_3$P=E) as precursors.[48] As consequence of the reduced reactivity the P=Se cleavage rate must be slowed, reducing the nucleation rate. For ZnSe nanocrystals, it was previously shown that the monomer concentration during the nucleation stage determines the formed phase, whereby at low monomer concentrations the WZ structure is favored.[49] In addition, Koster *et al.* calculated by density functional theory that the acetate ligands determine the crystal structure of CdSe nanoplatelets.[50] The lowest energy of acetate covered facets are found for the ZB-(001) ones and consequently the ZB structure is favored by acetate ligands. Moreover, they showed that the adsorption of neighboring acetate ions is energetically more unfavorable for WZ-(0001) facets than for ZB-(001) facets due the smaller surface area per acetate. As mentioned before, in WZ-Cd NSs the carboxylate surface ligands are partially replaced by Br$^-$ and Br-Hep. We suggest that the combination of a reduced nucleation rate as well the replacement of carboxylate ligands on the surface lead to the phase change in our system with increasing amount of bromine additive. As mentioned above, at high concentrations of Br-Hep CdBr$_2$ is isolated instead of CdSe NSs, because in that case the high amount of hydrobromic acid, which is in equilibrium with Br$^-$, must acidify the reaction solution and dissolve the formed CdSe NSs.

In conclusion, we report the successful size, shape and phase control of free-standing, ultrathin CdSe nanosheets using haloalkanes in a hot-injection synthesis. With the new and simple approach it is possible to tune the shape from sexangular to quadrangular to triangular and the phase from zinc blende to wurtzite. The shape and the phase changes are induced by *in-situ* formed halide ions, which alter nucleation and growth by acting as complexing agents and surface ligands. Due to the specific optical fingerprint such materials might be interesting for comparing spectroscopic studies and for applications in labeling and lighting devices.

ASSOCIATED CONTENT

Supporting Information Available:

Experimental details, quantum yield determination, XPS data, XRD data, optical spectroscopy data, EI mass spectra, TEM images, EDX and SAED data are provided. This material is available free of charge via the internet at http://pubs.acs.org.

AUTHOR INFORMATION

Corresponding Author: * E-Mail: klinke@chemie.uni-hamburg.de

Notes: The authors declare no competing financial interest.

ACKNOWLEDGMENT

F. G. and C.K. thank the European Research Council (Seventh Framework Program FP7, Project: ERC Starting Grant 2D-SYNETRA, 304980) for funding. C. K. acknowledges the German Research Foundation DFG for a Heisenberg scholarship KL 1453/9-2. B. H. J. thanks for funding in the frame of the following projects: S2013/MIT-2740 from Comunidad de Madrid, MAT2013-47395-C4-3-R and FIS2015-67367-C2-1-P from the Spanish Ministry of Economy and Competitiveness and CEAL-AL/2015-15 from UAM-Banco Santander. Further, the authors thank María Acebrón for experimental help. F. G. also thanks Frederic Leuffert and Mark-Tilo Schmitt for the CdSe synthesis with different bromine compounds.

# References

1. Butler, S. Z.; Hollen, S. M.; Cao, L.; Cui, Y.; Gupta, J. A.; Gutiérrez, H. R.; Heinz, T. F.; Hong, S. S.; Huang, J.; Ismach, A. F.; Johnston-Halperin, E.; Kuno, M.; Plashnitsa, V. V.; Robinson, R. D.; Ruoff, R. S.; Salahuddin, S.; Shan, J.; Shi, L.; Spencer, M. G.; Terrones, M.; Windl, W.; Goldberger, J. E. *ACS Nano* **2013**, *7*, 2898–2926.
2. Wang, Q. H.; Kalantar-Zadeh, K.; Kis, A.; Coleman, J. N.; Strano, M. S. *Nat. Nano.* **2012**, *7*, 699–712.
3. Xu, M.; Liang, T.; Shi, M.; Chen, H. *Chem. Rev.* **2013**, *113*, 3766–3798.
4. Novoselov, K. S.; Geim, A. K.; Morozov, S. V.; Jiang, D.; Zhang, Y.; Dubonos, S. V.; Grigorieva, I. V.; Firsov, A. A. *Science* **2004**, *306*, 666–669.
5. Rao, C. N. R.; Sood, A. K.; Subrahmanyam, K. S.; Govindaraj, A. *Angew. Chem. Int. Ed.* **2009**, *48*, 7752–7777.
6. Xiong, Y.; McLellan, J. M.; Chen, J.; Yin, Y.; Li, Z.-Y.; Xia, Y. *J. Am. Chem. Soc.* **2005**, *127*, 17118–17127.
7. Cao, Y. C. *J. Am. Chem. Soc.* **2004**, *126*, 7456–7457.
8. Fukuda, K.; Ebina, Y.; Shibata, T.; Aizawa, T.; Nakai, I.; Sasaki, T. *J. Am. Chem. Soc.* **2007**, *129*, 202–209.
9. Lv, R.; Robinson, J. A.; Schaak, R. E.; Du Sun; Sun, Y.; Mallouk, T. E.; Terrones, M. *Acc. Chem. Res.* **2015**, *48*, 56–64.
10. Aretouli, K. E.; Tsipas, P.; Tsoutsou, D.; Marquez-Velasco, J.; Xenogiannopoulou, E.; Giamini, S. A.; Vassalou, E.; Kelaidis, N.; Dimoulas, A. *Appl. Phys. Lett.* **2015**, *106*, 143105.
11. Coleman, J. N.; Lotya, M.; O'Neill, A.; Bergin, S. D.; King, P. J.; Khan, U.; Young, K.; Gaucher, A.; De, S.; Smith, R. J.; Shvets, I. V.; Arora, S. K.; Stanton, G.; Kim, H.-Y.; Lee, K.; Kim, G. T.; Duesberg, G. S.; Hallam, T.; Boland, J. J.; Wang, J. J.; Donegan, J. F.; Grunlan, J. C.; Moriarty, G.; Shmeliov, A.; Nicholls, R. J.; Perkins, J. M.; Grieveson, E. M.; Theuwissen, K.; McComb, D. W.; Nellist, P. D.; Nicolosi, V. *Science* **2011**, *331*, 568–571.
12. Lee, Y.-H.; Zhang, X.-Q.; Zhang, W.; Chang, M.-T.; Lin, C.-T.; Chang, K.-D.; Yu, Y.-C.; Wang, J. T.-W.; Chang, C.-S.; Li, L.-J.; Lin, T.-W. *Adv. Mater.* **2012**, *24*, 2320–2325.
13. Lauth, J.; Gorris, F. E. S.; Samadi Khoshkhoo, M.; Chassé, T.; Friedrich, W.; Lebedeva, V.; Meyer, A.; Klinke, C.; Kornowski, A.; Scheele, M.; Weller, H. *Chem. Mater.* **2016**, *28*, 1728–1736.
14. Li, L.; Chen, Z.; Hu, Y.; Wang, X.; Zhang, T.; Chen, W.; Wang, Q. *J. Am. Chem. Soc.* **2013**, *135*, 1213–1216.
15. Schliehe, C.; Juarez, B. H.; Pelletier, M.; Jander, S.; Greshnykh, D.; Nagel, M.; Meyer, A.; Foerster, S.; Kornowski, A.; Klinke, C.; Weller, H. *Science* **2010**, *329*, 550–553.
16. van der Stam, W.; Akkerman, Q. A.; Ke, X.; van Huis, M. A.; Bals, S.; Mello Donega, C. de. *Chem. Mater.* **2015**, *27*, 283–291.
17. Jun, Y.-w.; Choi, J.-s.; Cheon, J. *Angew. Chem. Int. Ed.* **2006**, *45*, 3414–3439.
18. Park, J.; Joo, J.; Kwon, S. G.; Jang, Y.; Hyeon, T. *Angew. Chem. Int. Ed.* **2007**, *46*, 4630–4660.
19. Ouyang, J.; Zaman, M. B.; Yan, F. J.; Johnston, D.; Li, G.; Wu, X.; Leek, D.; Ratcliffe, C. I.; Ripmeester, J. A.; Yu, K. *J. Phys. Chem. C* **2008**, *112*, 13805–13811.
20. Ithurria, S.; Dubertret, B. *J. Am. Chem. Soc.* **2008**, *130*, 16504–16505.
21. Ithurria, S.; Bousquet, G.; Dubertret, B. *J. Am. Chem. Soc.* **2011**, *133*, 3070–3077.
22. Bouet, C.; Mahler, B.; Nadal, B.; Abecassis, B.; Tessier, M. D.; Ithurria, S.; Xu, X.; Dubertret, B. *Chem. Mater.* **2013**, *25*, 639–645.

23. Joo, J.; Son, J. S.; Kwon, S. G.; Yu, J. H.; Hyeon, T. *J. Am. Chem. Soc.* **2006**, *128*, 5632–5633.

24. Son, J. S.; Wen, X.-D.; Joo, J.; Chae, J.; Baek, S.-i.; Park, K.; Kim, J. H.; An, K.; Yu, J. H.; Kwon, S. G.; Choi, S.-H.; Wang, Z.; Kim, Y.-W.; Kuk, Y.; Hoffmann, R.; Hyeon, T. *Angew. Chem. Int. Ed.* **2009**, *48*, 6861–6864.

25. Yang, J.; Son, J. S.; Yu, J. H.; Joo, J.; Hyeon, T. *Chem. Mater.* **2013**, *25*, 1190–1198.

26. Ithurria, S.; Tessier, M. D.; Mahler, B.; Lobo, R. P. S. M.; Dubertret, B.; Efros, A. L. *Nat. Mater.* **2011**, *10*, 936–941.

27. Pelton, M.; Ithurria, S.; Schaller, R. D.; Dolzhnikov, D. S.; Talapin, D. V. *Nano Lett.* **2012**, *12*, 6158–6163.

28. Achtstein, A. W.; Schliwa, A.; Prudnikau, A.; Hardzei, M.; Artemyev, M. V.; Thomsen, C.; Woggon, U. *Nano Lett.* **2012**, *12*, 3151–3157.

29. Mahler, B.; Nadal, B.; Bouet, C.; Patriarche, G.; Dubertret, B. *J. Am. Chem. Soc.* **2012**, *134*, 18591–18598.

30. Li, Z.; Peng, X. *J. Am. Chem. Soc.* **2011**, *133*, 6578–6586.

31. Chen, D.; Gao, Y.; Chen, Y.; Ren, Y.; Peng, X. *Nano Lett* **2015**, *15*, 4477–4482.

32. Scott, R.; Achtstein, A. W.; Prudnikau, A. V.; Antanovich, A.; Siebbeles, L. D. A.; Artemyev, M.; Woggon, U. *Nano Lett* **2016**, *16*, 6576–6583.

33. Fox, M. *Optical Properties of Solids*, 2. edition; Oxford University Press: Oxford, New York, 2010.

34. Gao, Y.; Peng, X. *J. Am. Chem. Soc.* **2014**, *136*, 6724–6732.

35. Feigl, C. A.; Barnard, A. S.; Russo, S. P. *Phys. Chem. Chem. Phys.* **2012**, *14*, 9871–9879.

36. Singh, V.; Sharma, P. K.; Chauhan, P. *Mater. Chem. Phys.* **2010**, *121*, 202–207.

37. Ricolleau, C.; Audinet, L.; Gandais, M.; Gacoin, T.; Boilot, J.-P.; Chamarro, M. *Proceedings of the seventh international conference on II-VI compounds and devices* **1996**, *159*, 861–866.

38. Lim, J.; Bae, W. K.; Park, K. U.; Zur Borg, L.; Zentel, R.; Lee, S.; Char, K. *Chem. Mater.* **2013**, *25*, 1443–1449.

39. Yu, W. W.; Wang, Y. A.; Peng, X. *Chem. Mater.* **2003**, *15*, 4300–4308.

40. Bielewicz, T.; Ramin Moayed, M. M.; Lebedeva, V.; Strelow, C.; Rieckmann, A.; Klinke, C. *Chem. Mater.* **2015**, *27*, 8248–8254.

41. Meyns, M.; Iacono, F.; Palencia, C.; Geweke, J.; Coderch, M. D.; Fittschen, U. E. A.; Gallego, J. M.; Otero, R.; Juárez, B. H.; Klinke, C. *Chem. Mater.* **2014**, *26*, 1813–1821.

42. To affirm this hypothesis of the in-situ formed Br⁻, we used cadmium bromide and different ammonium bromide compounds as direct bromide source. In all cases, the synthesis failed due to the insolubility of the bromide salts in the solvent diphenyl ether.

43. Seals, R. D.; Alexander, R.; Taylor, L. T.; Dillard, J. G. *Inorg. Chem.* **1973**, *12*, 2485–2487.

44. Briggs, D.; Seah, M. P. *Practical Surface Analysis. Volume 1 Auger and X-ray Photoelectron Spectroscopy*, 2. edition, reprint; John Wiley & Sons: Chichester, 1994.

45. Zhou, X.-L.; Solymosi, F.; Blass, P. M.; Cannon, K. C.; White, J. M. *Surf. Sci.* **1989**, *219*, 294–316.

46. Mazov, I.; Krasnikov, D.; Stadnichenko, A.; Kuznetsov, V.; Romanenko, A.; Anikeeva, O.; Tkachev, E. *J. Nanotechnol.* **2012**, *2012*, 5.

47. Clayden, J.; Greeves, N.; Warren, S.; Wothers, P. *Organic Chemistry;* Oxford University Press: Oxford, 2001.

48. Steckel, J. S.; Yen, B. K. H.; Oertel, D. C.; Bawendi, M. G. *J. Am. Chem. Soc.* **2006**, *128*, 13032–13033.

49. Omata, T.; Uesugi, H.; Kita, M. *Journal of Crystal Growth* **2014**, *394*, 81–88.

50. Koster, R. S.; Fang, C.; van Blaaderen, A.; Dijkstra, M.; van Huis, M. A. *Phys. Chem. Chem. Phys.* **2016,** *18*, 22021–22024.

**Figures**

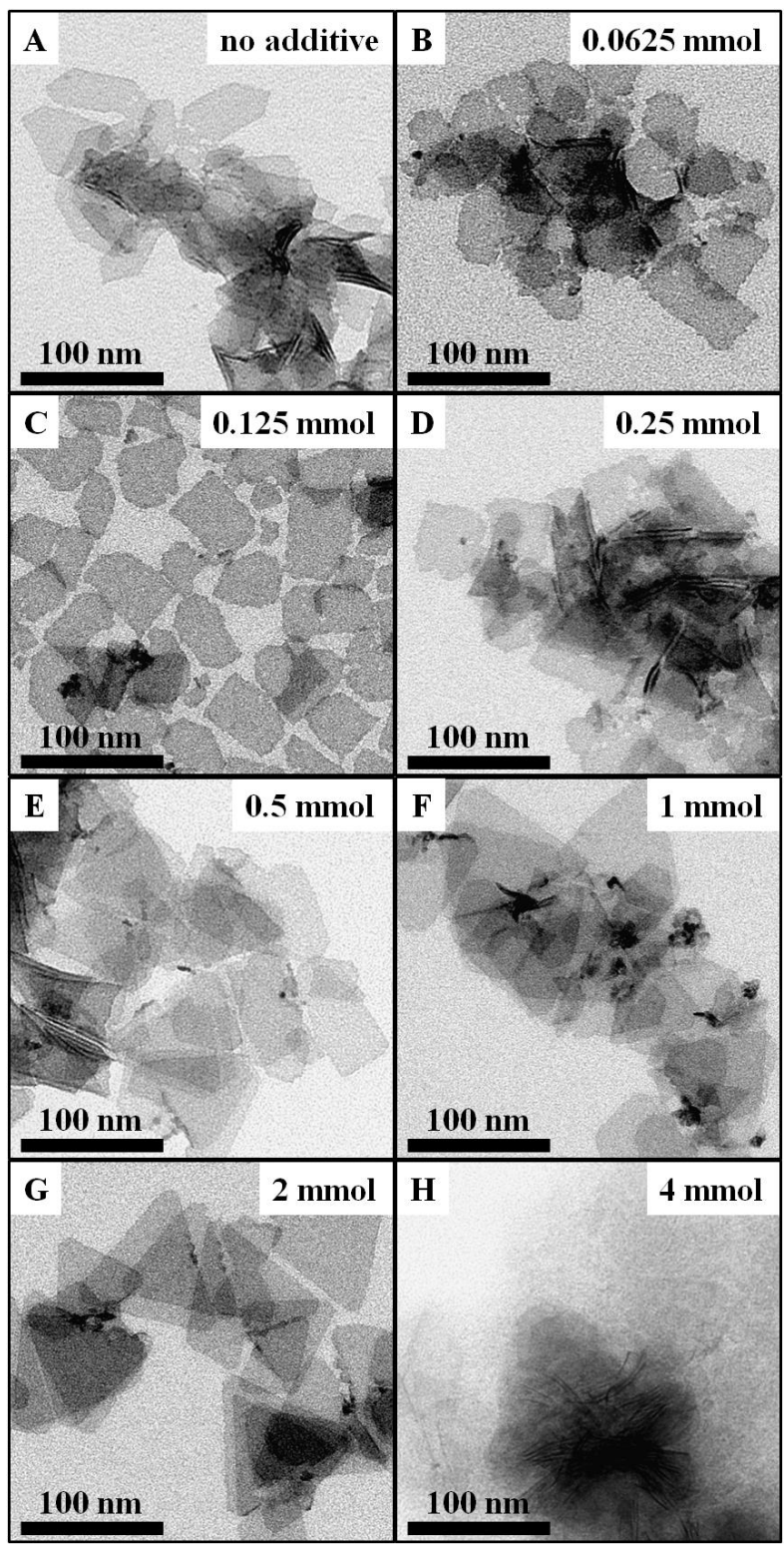

**Figure 1.** TEM images of ultrathin CdSe NSs synthesized with different amounts of Br-Hep. (A) Sexangular (no additive), (B, C, D) quadrangular (0.0625, 0.125 and 0.25 mmol Br-Hep), (E) mixture-shaped (0.5 mmol Br-Hep), (F) triangular (1 mmol Br-Hep), (G) triangular (2 mmol Br-Hep), and (H) undefined NSs (4 mmol Br-Hep).

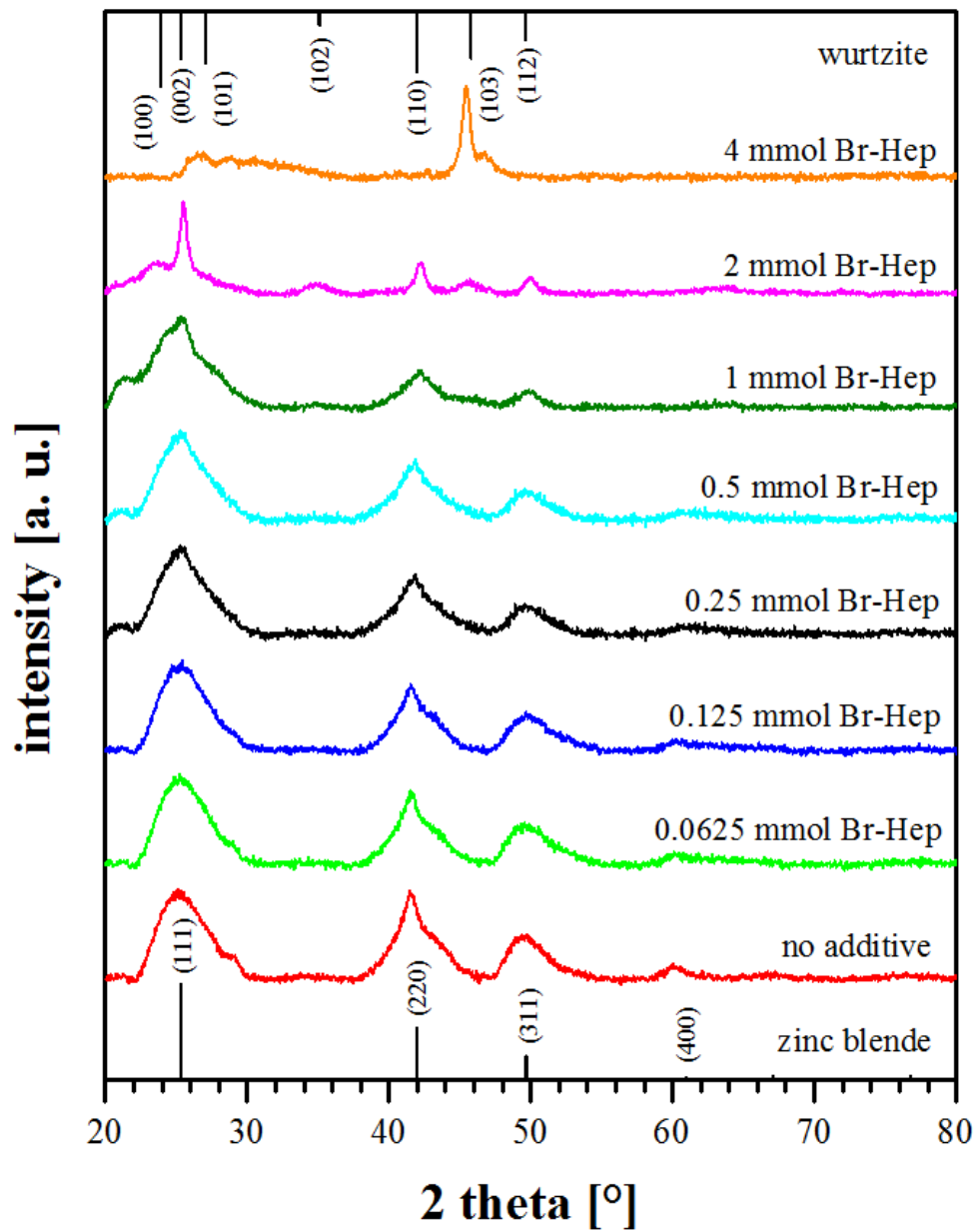

**Figure 2.** XRD of ultrathin CdSe NSs synthesized with different amounts of Br-Hep. At the top (wurtzite, ICPDS 00-008-0459) and the bottom (zinc blende, ICPDS 00-019-0191) the diffractograms of the bulk CdSe is shown. The peaks of the sample with 4 mmol Br-Hep show a strong texture effect and are assigned to the bulk CdBr$_2$ (ICPDS 00-010-0438).

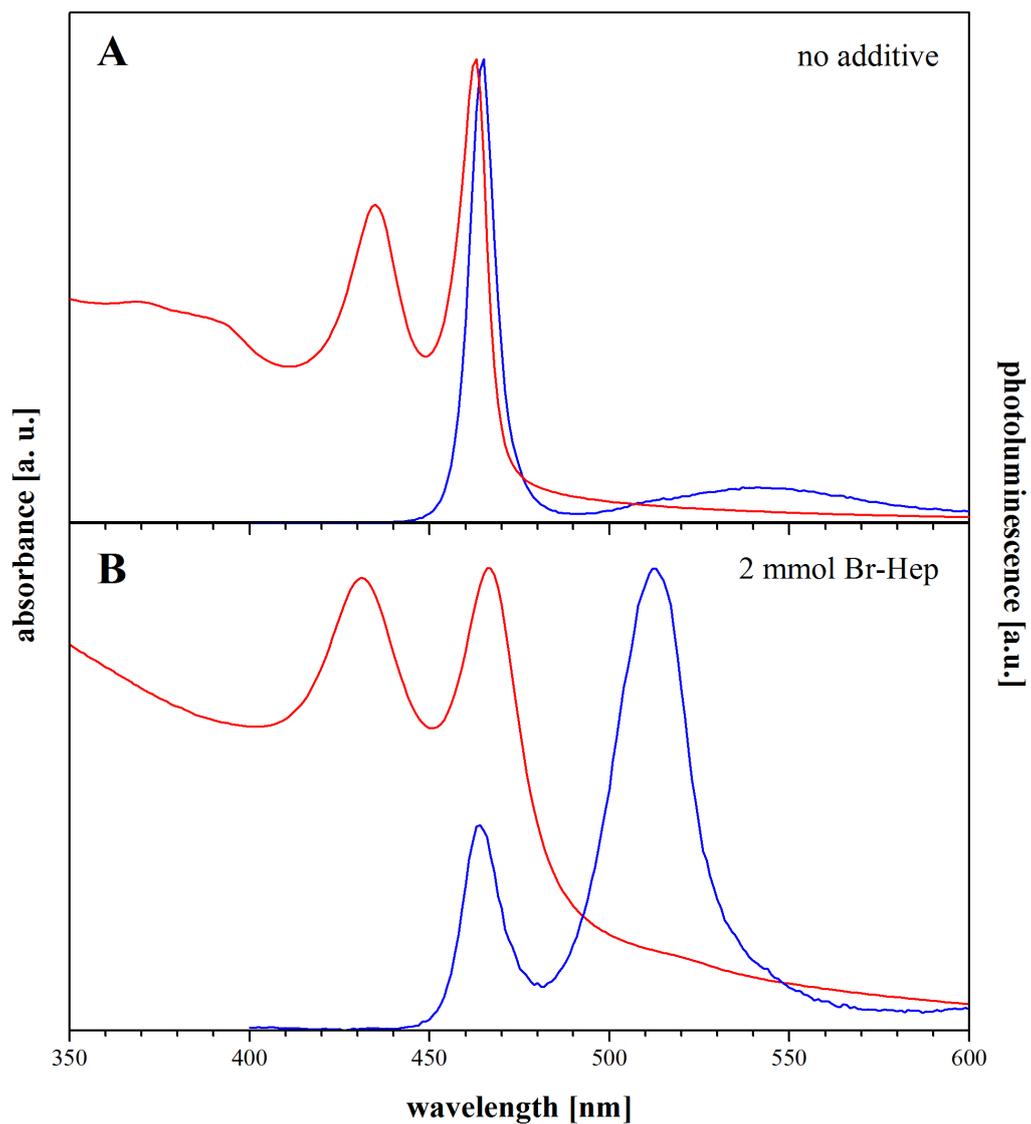

**Figure 3.** Absorption (blue line) and photoluminescence (red line) spectra of phase-different CdSe NSs. (A) ZB-CdSe NSs synthesized without Br-Hep and (B) WZ-CdSe NSs synthesized with 2 mmol Br-Hep.

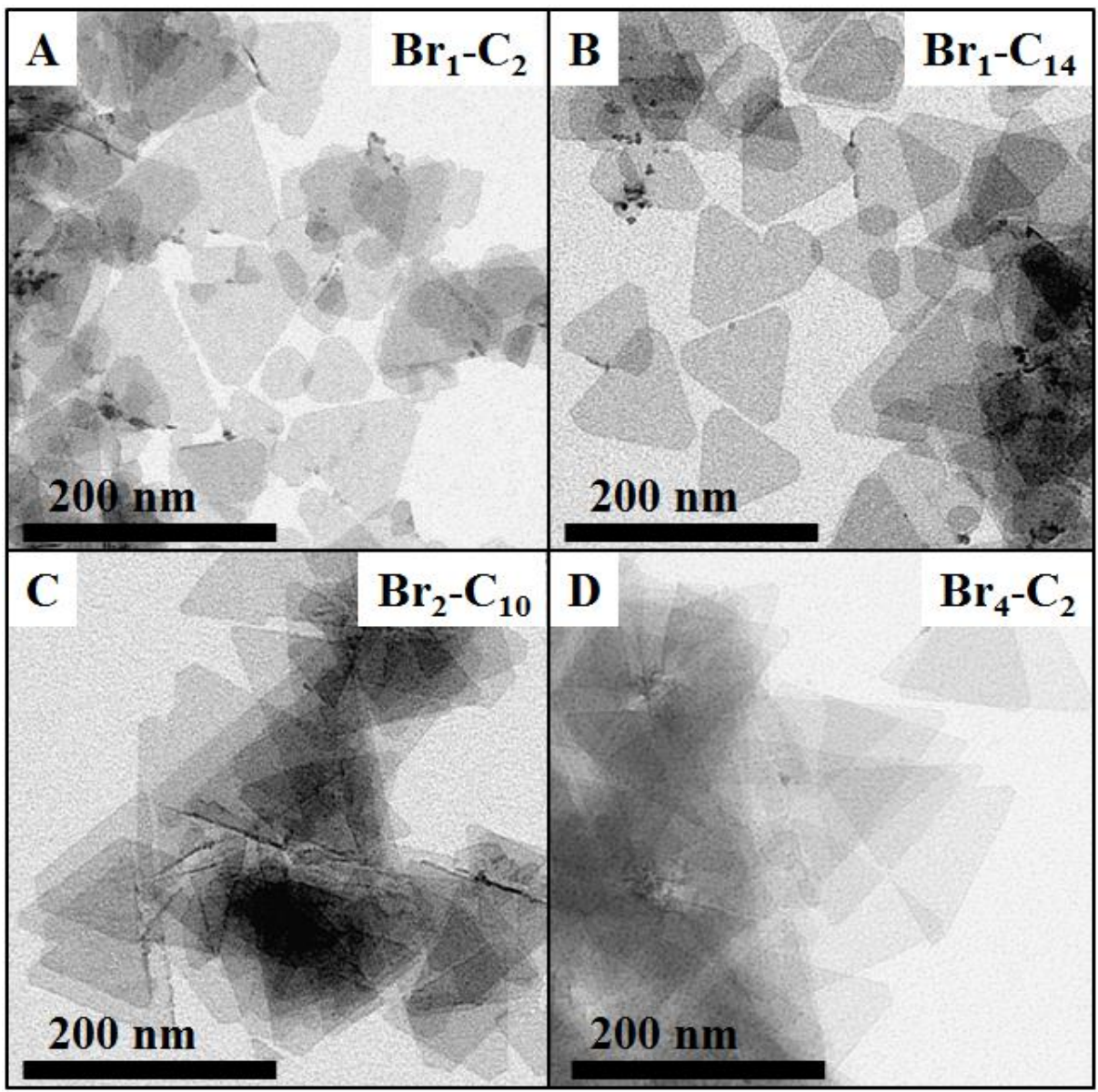

**Figure 4.** TEM images of ultrathin CdSe NSs synthesized with different bromoalkanes to show the influence of chain length and number of bromine atoms on the synthesis. (A) 1-bromoethane ($Br_1$-$C_2$, 1 mmol), (B) 1-bromotetradecane ($Br_1$-$C_{14}$, 0.5 mmol), (C) 1,10-dibromodecane ($Br_2$-$C_{10}$, 0.25 mmol), and (D) 1,1,2,2-tetrabromoethane ($Br_4$-$C_2$, 0.125 mmol).

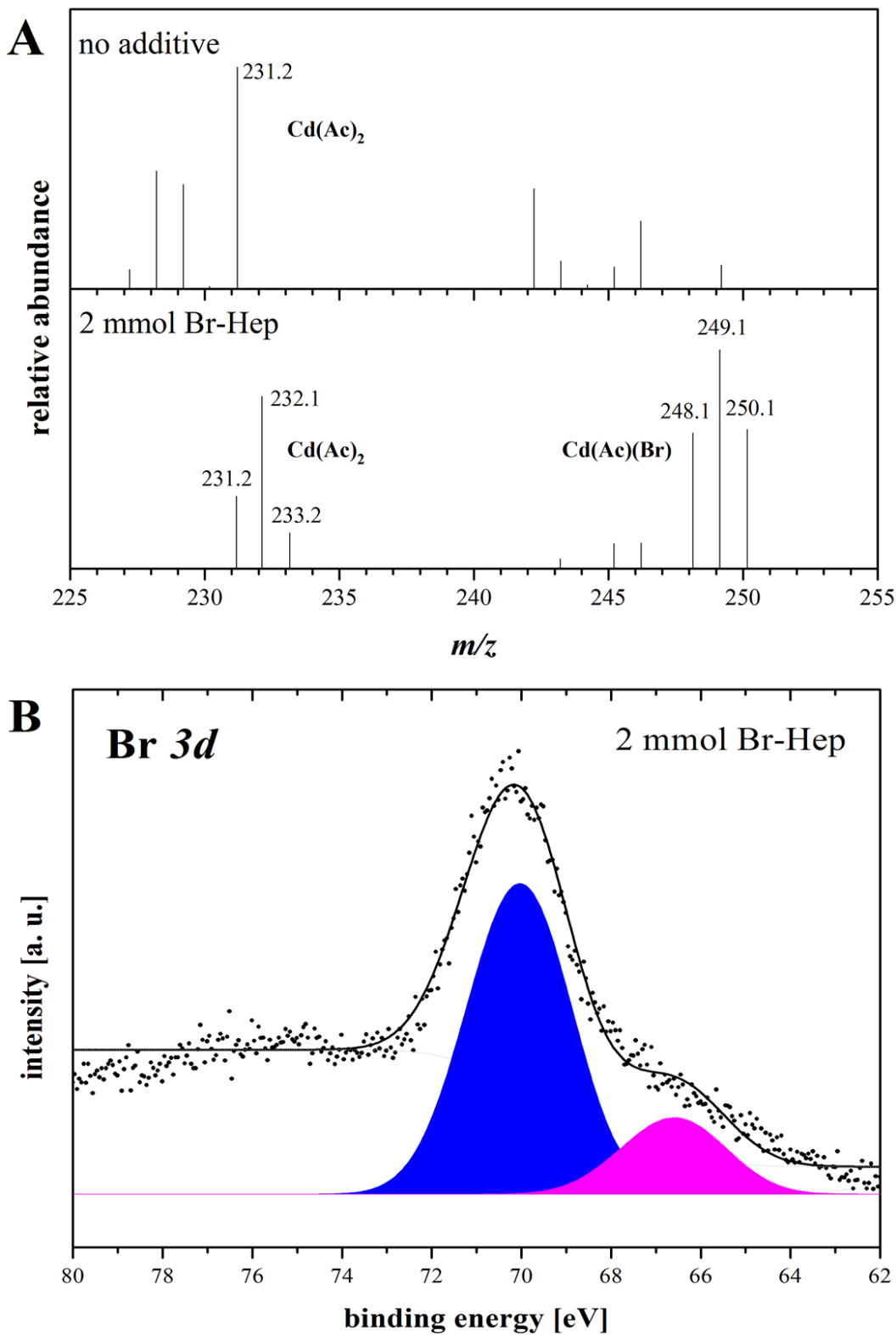

**Figure 5.** (A) EI mass spectra of the reaction solution after the vacuum step at 80 °C (no additive, before Br-Hep injection), and directly before the TOPSe injection at 180 °C (2 mmol Br-Hep, 10 min after Br-Hep-Injection) in the range from *m/z* 225 to 255. (B) XPS spectrum of the region of Br *3d* of CdSe NSs synthesized with 2 mmol of Br-Hep. The XPS spectrum was obtained at a photon energy of 1487 eV.

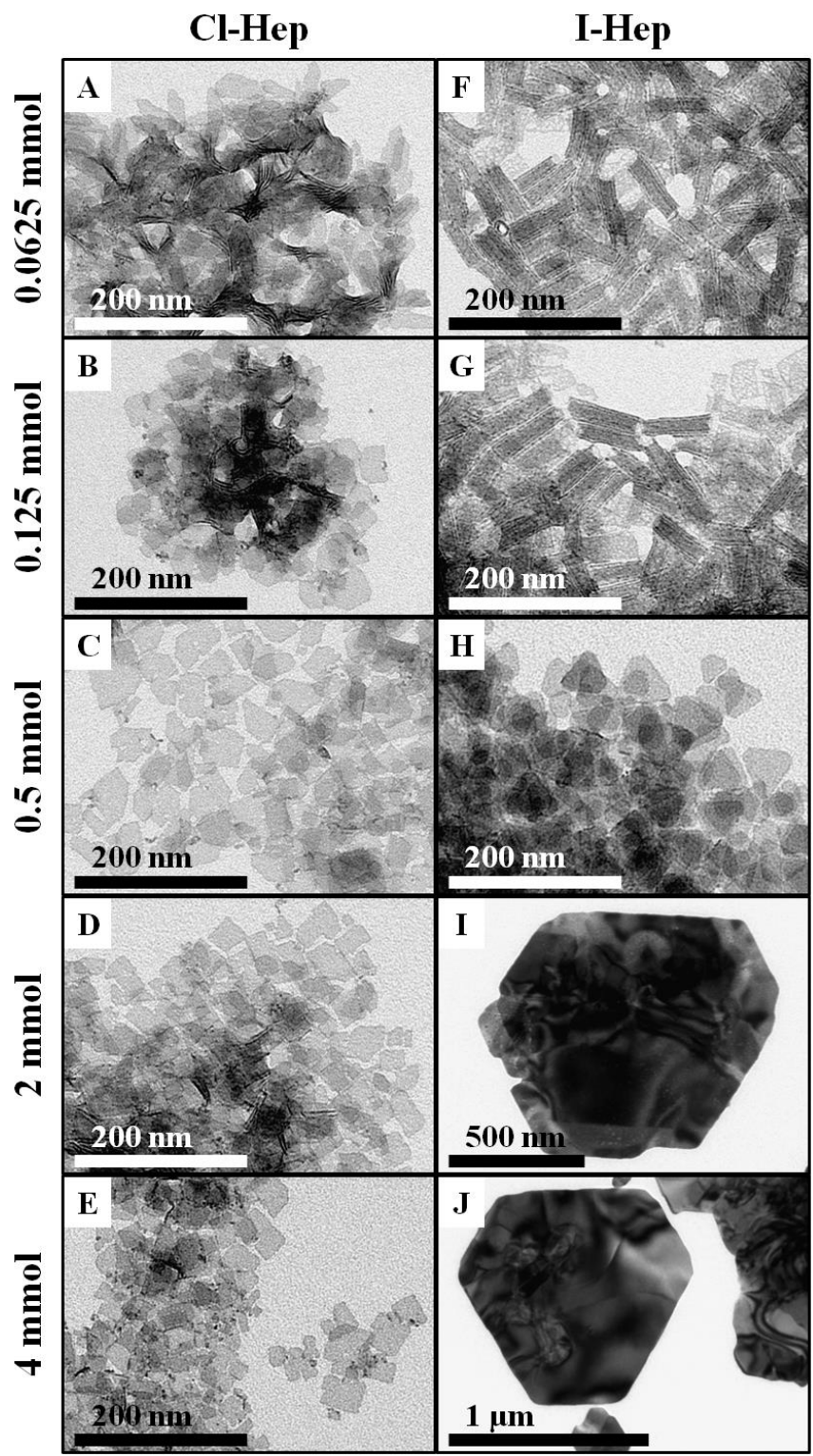

**Figure 6.** TEM images of ultrathin CdSe NSs synthesized with different amounts of Cl-Hep (A-E) and I-Hep (F-J). The addition of Cl-Hep accompanies with a shape change from sexangular (A) to quadrangular (B-E) CdSe NSs, while the crystal phase is kept at the ZB structure. Increasing amounts of I-Hep lead to ZB-CdSe rolled-up NSs (F, G) to WZ-CdSe triangular NSs (H) to CdI$_2$ hexagons (I, J).

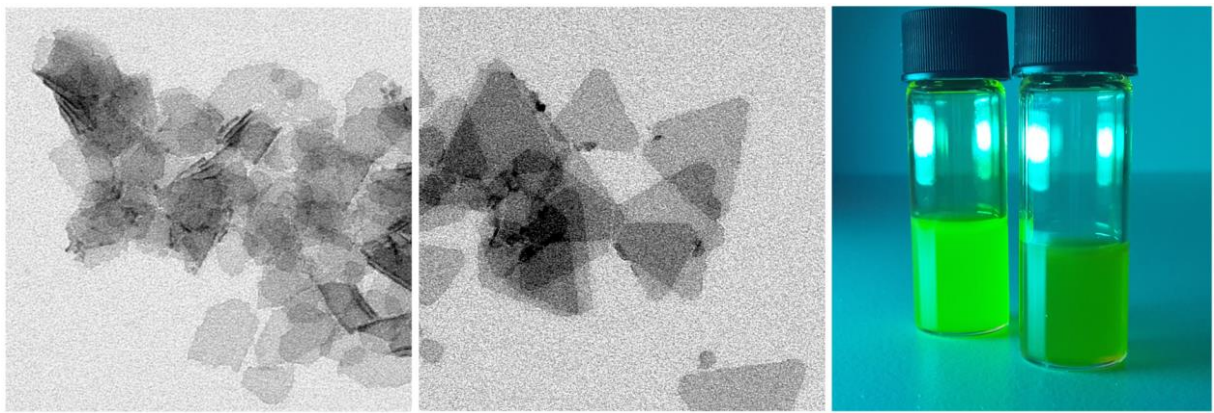

**Figure Table of Content**